\documentclass[a4paper,11pt]{cpc-hepnp}

\usepackage{textcomp} 
\usepackage{slashed}
\usepackage{epsfig,latexsym,cancel,amssymb,amsmath,verbatim,mathrsfs}
\usepackage{color}
\usepackage{graphicx}
\usepackage{enumitem}
\usepackage[colorlinks,citecolor=blue]{hyperref}

\newcommand{\beq}{\begin{equation}}
\newcommand{\eeq}{\end{equation}}
\newcommand{\bea}{\begin{eqnarray}}
\newcommand{\eea}{\end{eqnarray}}

\newcommand{\GeV}{\mathrm{GeV}}

\allowdisplaybreaks[1]

\begin{document}

\fancyhead[c]{\small ~~~~~~~~~~~~~~~~~~~~~~~~~~~~~~~~~~~~~~~~~~
	Prepared for Chinese Physics C~~~~~~~~~~~~~~~~~~~~~~~~~~~~~~
	~~~~~~~~~~~~~~~ LA-UR-21-31766}

\title{Searching for the axion-like particle at the EIC}

\author{Yandong Liu$^{1,2}$\email{ydliu@bnu.edu.cn}, %
	\quad Bin Yan$^{3,4}$\email{yanbin@ihep.ac.cn (corresponding author)}
}

\maketitle

\address{%
$^1$Key Laboratory of Beam Technology of Ministry of Education, \\College of Nuclear Science and Technology, Beijing Normal University, Beijing 100875, China\\	
$^2$Beijing Radiation Center, Beijing 100875, China\\
$^3$Institute of High Energy Physics, Chinese Academy of Sciences, Beijing 100049, China\\
$^4$Theoretical Division, Group T-2, MS B283, Los Alamos National Laboratory, P.O. Box 1663, Los Alamos, NM 87545, USA
}

\begin{abstract}
The axion-like particle (ALP) is a well motivated new particle candidate of beyond the Standard Model. In this work, we propose to probe the ALP  through the photon fusion scattering at the upcoming Electron-Ion Collider (EIC) with electron and proton energy $E_e=20~{\rm GeV}$ and $E_p=250~{\rm GeV}$.  It shows that we could constrain the effective coupling strength between ALP and photons to  be $0.2~{\rm TeV}^{-1}$ at $2\sigma$ confidence level with the integrated luminosity of $300~{\rm fb}^{-1}$ for the mass range $m_a\in [5,40]~{\rm GeV}$. Such bound could be much improved if we consider the nucleus beam at the EIC. We also demonstrate that the limits from the EIC could be stronger than the off $Z$-pole measurement at the LEP and the Light-by-Light scattering with pp collision at the LHC. 
\end{abstract}

\maketitle

\section{Introduction}
\label{sec:intr}
The axion-like particles (ALPs) are widely predicted in many new physics beyond the Standard Model (BSM). A well know example is that the Pseudo-Nambu-Goldstone boson (PNGB) from the new global symmetry breaking, which was designed to solve the strong CP problem~\cite{Peccei:1977hh,Peccei:1977ur}.  The ALP has received much attention in the particle physics  and cosmology communities because it could solve the naturalness problems~\cite{Freese:1990rb,Graham:2015cka} and also  could be a compelling dark matter candidate of the universe~\cite{Preskill:1982cy,Abbott:1982af,Dine:1982ah}. The landscape of ALPs is very rich and the  phenomenology is determined by their mass and the couplings with the SM particles.  In general, the ALPs could couple to gauge bosons,  fermions and Higgs boson (see in Refs.~\cite{Chala:2020wvs,Choi:2020rgn,Bauer:2021mvw,Galda:2021hbr,Bauer:2020jbp,Bonilla:2021ufe} for a general discussion). Many dedicated experiments have been proposed to search the ALPs based on their mass region and couplings. For example, the interactions between ALPs and fermions can be probed in rare decays~\cite{Cornella:2019uxs,Dolan:2014ska,Calibbi:2020jvd,Carmona:2021seb,Ma:2021jkp,Bauer:2021mvw,Cheung:2021mol,Chakraborty:2021wda,Bertholet:2021hjl}. The bound of the couplings to the gauge bosons could be constrained by the astrophysics and cosmology observations, e.g. the stellar evolution, big bang nucleosynthesis, anisotropies in the cosmic microwave background, etc for a light ALP~\cite{Cadamuro:2011fd,Payez:2014xsa,Millea:2015qra,Jaeckel:2017tud} and by collider searches for a heavy ALP with a Light-by-Light or vector-boson fusion type scattering~\cite{OPAL:2002vhf,Jaeckel:2015jla,Knapen:2016moh,Bauer:2017ris,Dolan:2017osp,CMS:2018erd,Yue:2019gbh,Belle-II:2020jti,Yang:2021xgt,Davoudiasl:2021haa,Brivio:2021fog,BuarqueFranzosi:2021kky,CMS:2021xor,ATLAS:2014jdv,ATLAS:2015rsn,Baldenegro:2018hng,Ren:2021prq,Wang:2021uyb,Inan:2020aal,Inan:2020kif,Zhang:2021sio,Steinberg:2021wbs,Florez:2021zoo,ATLAS:2017fur}.

\begin{figure}
\centering
\includegraphics[scale=0.4]{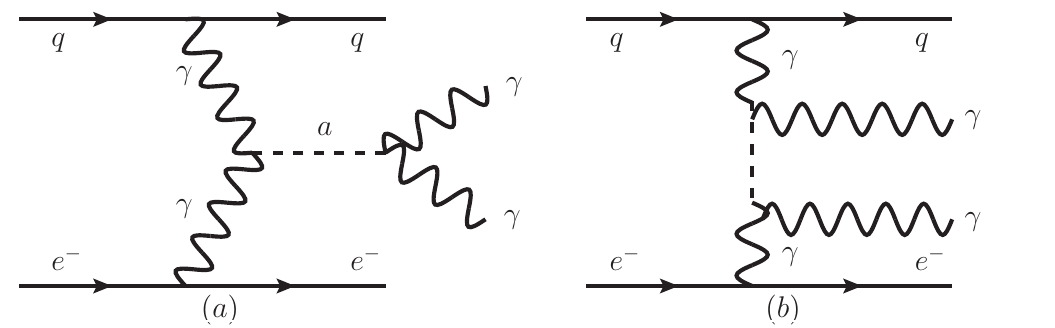}
\caption{Feynman diagrams of the parton process $e^-q\to e^- j \gamma\gamma$. } \label{FIG:ep2ejaa}
\end{figure}

In this work, we will consider the case that the ALP $a$ predominantly couples to photons, so that the branching ratio ${\rm BR}(a\to \gamma\gamma)=1$ and the effective Lagrangian can be parameterized as,
\beq \label{Eq:lag}
\mathcal{L}_{\rm eff}=\frac{1}{2}\left(\partial_\mu a\right)^2-\frac{1}{2}m_a^2a^2-g_{a\gamma\gamma}aF_{\mu\nu}\widetilde{F}^{\mu\nu},
\eeq
where $a$ is the ALP with mass $m_a$ and $(\widetilde{F}_{\mu\nu}) ~F_{\mu\nu}$ is the (dual) field strength tensor of the photon. The coupling strength $g_{a\gamma\gamma}$ has be seriously constrained by the electron/proton beam-dump experiments~\cite{NA64:2020qwq,Dobrich:2015jyk}, $e^+e^-\to \gamma$+invisible~\cite{Dolan:2017osp}, inclusive $e^+e^-\to \gamma\gamma$~\cite{OPAL:2002vhf}, a photon-beam experiment~\cite{Aloni:2019ruo} and $e^+e^-\to \gamma a (\to \gamma\gamma)$ at the Belle II~\cite{Belle-II:2020jti} with mass range from MeV to 10 GeV. Above the 10 GeV,  the parameter space can be probed by the Light-by-Light scattering in heavy-ion collisions at the Large Hadron Collider (LHC) energy~\cite{ATLAS:2017fur,CMS:2018erd}, $e^+e^-\to \gamma a (\to \gamma\gamma)$ at the LEP~\cite{Jaeckel:2015jla}, the measurements at the LHC~\cite{ATLAS:2014jdv,ATLAS:2015rsn,Baldenegro:2018hng,Ren:2021prq,Wang:2021uyb,Florez:2021zoo} and future lepton colliders~\cite{Inan:2020aal,Inan:2020kif,Zhang:2021sio,Steinberg:2021wbs}. However, the bound from GeV to tens of GeV is less limited by the current experiments compared to the MeV (or smaller) mass range. 
Such light ALP is very attractive for the community because it could be a natural product from a new global symmetry breaking if it is a PNGB. 
In this paper, we propose to search the ALP through the photon fusion production at the  upcoming Electron-Ion Collider (EIC) (see Fig.~\ref{FIG:ep2ejaa})~\footnote{The elastic scattering of two photons could also be used to search the ALP, but the signature is different from our scenario~\cite{Davoudiasl:2021mjy}.}, which could be complementary to the measurements at the LHC, heavy-ion  collisions at the LHC energy and lepton colliders for probing the ALP in this mass range.

\section{The production of ALP}
The ALP could be produced through $s$-channel and $t$-channel photon fusion scattering at the EIC~\footnote{The gauge-invariant operators can generate the effective couplings $a\gamma\gamma$, $aZ\gamma$ and $aZZ$ at the same time, but the contributions from $aZ\gamma$ and $aZZ$ will be highly suppressed by the propagator of $Z$ boson and can be ignored.}; see Fig.~\ref{FIG:ep2ejaa}. The production rate of the signal depends on the coupling strength $g_{a\gamma\gamma}$, ALP mass $m_a$ and branching ratio ${\rm BR}(a\to\gamma\gamma)$, i.e.
\beq
\sigma(e^-p\to e^-\gamma\gamma j)=g_{a\gamma\gamma}^2\sigma_s(m_a)\times {\rm BR}(a\to\gamma\gamma)+g_{a\gamma\gamma}^4\sigma_t(m_a),
\eeq
where $\sigma_{s,t}$ denotes the cross section from the $s$-channel and $t$-channel, respectively. The narrow width approximation has been applied for the $s$-channel scattering since the decay width $\Gamma_a\ll m_a$~\cite{HPILUHNL:NWA}. 
We checked the interference effects between the signal and background is negligible and can be ignored in our analysis.  For simplicity,  we take the branching ratio ${\rm BR}(a\to\gamma\gamma)=1$ in this work.

\begin{figure}
\centering
\includegraphics[scale=0.55]{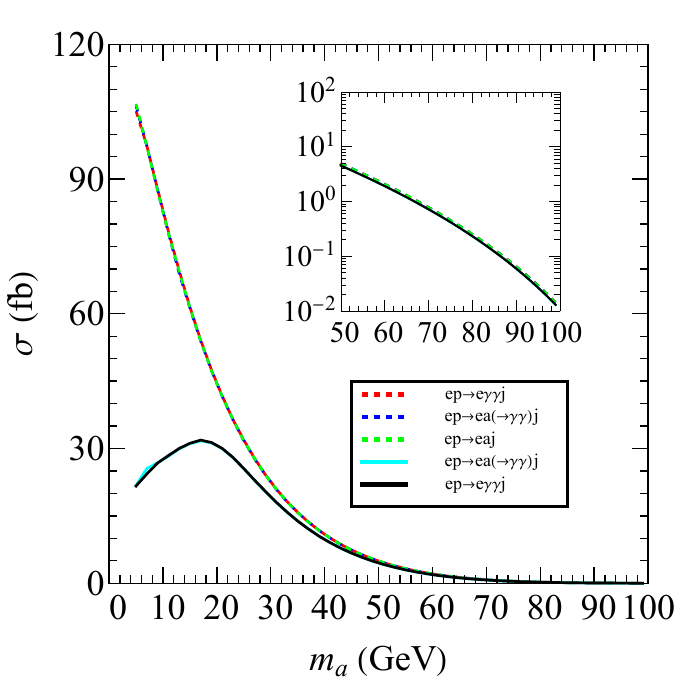}
\caption{The cross section of the signal process $e^-p\rightarrow e^- \gamma \gamma j$ as a function of the ALP mass with the coupling strength $g_{a\gamma\gamma}=1~{\rm TeV}^{-1}$. The solid and dashed lines are corresponding to the production rate with and without the kinematic cuts (see Eq.~\eqref{EQ:bcuts}) for the photons in the final states, respectively. We consider both the $s$-channel and $t$-channel for the red line, while only the contribution from $s$-channel has been considered for the green and blue lines.}
\label{FIG:cs}
\end{figure}

Below, we utilize the MadGraph5~\cite{Alwall:2014hca} to calculate the signal cross section with electron and proton energy $E_e=20~{\rm GeV}$ and $E_p=250~{\rm GeV}$ at the leading order
with NNPDF sets~\cite{Ball:2012cx}. The factorization and renormalization scales in our calculation have been chosen as the default scale in the MadGraph5, i.e. the transverse mass. To avoid the soft and collinear divergence, the following kinematic cuts for the final states have been applied,
 \begin{align} 
&p_T^{j,e,\gamma} > 5~ \GeV,~|\eta^{j,e,\gamma}|<5, \nonumber \\
&\Delta R(e,j)>0.4,~\Delta R(\gamma,\gamma)>0.4, \nonumber \\
&\Delta R(e,\gamma)>0.4, ~\Delta R(j,\gamma)>0.4, \label{EQ:bcuts}
\end{align}
where $p_T^m$ and $\eta^m$ with $m=j,e,\gamma$ denotes the transverse momentum and pseudorapidity of particle $m$, respectively. The cone distance $\Delta R(m,n)=\sqrt{(\eta^m-\eta^n)^2+(\phi^m-\phi^n)^2}$ with $\phi^m$ denoting the azimuthal angle of particle $m$.
Figure~\ref{FIG:cs} displays the production rates for the processes $e^-p\to e^- \gamma\gamma j$ (red dashed line), $e^-p\to e^-a(\to\gamma\gamma)j$ (blue dashed line) and $e^-p\to e^-aj$ (green dashed line) without any kinematic cuts for the photons in the final state.  It clearly shows that the cross section of the signal is dominantly determined by the $s$-channel scattering and the contribution from $t$-channel is negligible, because of the suppression of the phase space.
In addition, the narrow width approximation for the $s$-channel production works very well in this process. In the same figure, the solid cyan and black lines show the cross sections from $e^-p\to e^- \gamma\gamma j$ and  $e^-p\to e^-a(\to\gamma\gamma)j$ after we impose the kinematic cuts in Eq.~\eqref{EQ:bcuts} for the photons, respectively.  We note that the kinematic cuts of the photons will decrease the cross section significantly when ALP mass $m_a<40~{\rm GeV}$. It arises from the fact that the invariant mass of photon pair from an on-shell ALP decay is given by $m_a^2=m_{\gamma\gamma}^2\simeq p_T^{\gamma 1}p_T^{\gamma 2}\Delta R(\gamma 1,\gamma 2)$, where $p_T^{\gamma 1,\gamma 2}$ denotes the transverse momentum of the photons. For a heavy ALP, the kinematic cuts could be satisfied automatically, while there is a phase space suppression effect for a light ALP which is induced by the kinematic cuts of the photons (see Eq.~\eqref{EQ:bcuts}). Such effect will also generate a peak around  $m_a\sim 16 ~{\rm GeV}$ for the production rate distributions of the processes with the cuts on the photons (see cyan and black lines).

\section{Collider Simulation}
Next we perform a detailed Monte Carlo simulation to explore the potential of probing the ALP at the EIC. The major  irreducible backgrounds come from the processes $e^-p\to e^- \gamma \gamma j$ and $e^-p\to e^- \gamma \gamma jj$. The cross sections after we including the kinematic cuts in Eq.~\eqref{EQ:bcuts} are $\sigma(e^-p\to e^- \gamma \gamma j)=46.9~{\rm fb}$ and $\sigma(e^-p\to e^- \gamma \gamma jj)=6.4~{\rm fb}$. 
We also take into account the possibility that an electron or jet is misidentified to be a photon in this study. The reducible backgrounds could be from the processes $e^-p\to e^-\gamma j ~(18.66~{\rm pb})$ and $e^-p\to e^-\gamma jj ~(2.22~{\rm pb})$. The numbers shown inside the bracket denote the production cross section after imposing the cuts in Eq.~\eqref{EQ:bcuts}. The other backgrounds involving multi-electrons and jets, e.g.  $e^-p\to e^- e^- e^+ j$ and $e^-p\to e^-jj(j)$ are negligible when we consider the basic cuts in Eq.~\eqref{EQ:bcuts} and the mistag efficiencies.   We generate both the signal and backgrounds by MadGraph5~\cite{Alwall:2014hca} with the kinematic cuts in Eq.~\eqref{EQ:bcuts}.  The parton level events are passed to the PYTHIA8~\cite{Sjostrand:2014zea} for parton showering and hadronization and the detector effects are simulated by Delphes~\cite{deFavereau:2013fsa}. In the detector simulation, we use the EIC delphes card which was generated by M. Arratia and S. Sekula~\cite{Arratia:2021uqr}, based on parameters in~\cite{AbdulKhalek:2021gbh}  and utilized in~\cite{Arratia:2020nxw,Arratia:2020azl,Cirigliano:2021img}. The detector parameters in Ref.~\cite{AbdulKhalek:2021gbh} have dictated the tracking momentum resolution, secondary-vertex resolutions, calorimeter energy resolutions, as well as particle identification performance at the EIC. 
Among of them, the photon energy resolution would play the key role for our simulation, which is $\delta E/E=\mathcal{A}/E/{\rm{GeV}}\oplus\mathcal{B}/\sqrt{E/{\rm{GeV}}}\oplus{\mathcal{C}}$ with $\mathcal{A}=1\%$, $\mathcal{B}=2.5\%$ and $\mathcal{C}=1\%$. Now, we further require a set of preselection cuts on the reconstruction objects as follows:
\begin{align}
n^\gamma \ge 2,~~ |\eta^m|<3.0, ~~ p_T^{m}>7 {~\rm{GeV}},~~ \Delta R(m,n)>0.4,
\label{eq:cut2}
\end{align}
where $m,n=\gamma,e,j$ denote the photon, electron and jet from the detector reconstruction, respectively.
Note that the kinematic threshold for the reconstruction objects at the EIC could be looser compared to the LHC due to the much lower collider energy; see the cuts of HERA as a reference~\cite{H1:2016goa}.

\begin{figure} 
\centering
\includegraphics[scale=0.7]{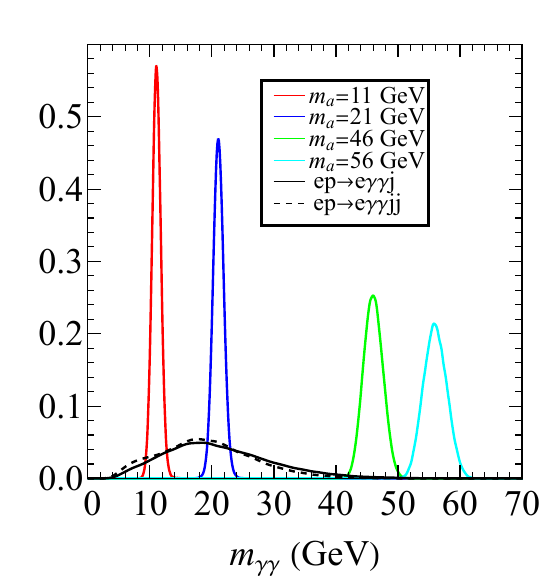}
\caption{The normalized invariant mass distribution of the photon pair from the signal process $e^-p\rightarrow e^- \gamma \gamma j$ and the SM backgrounds $e^-p\rightarrow e^- \gamma \gamma j$, $e^-p\rightarrow e^- \gamma \gamma j j$ and $e^-p\rightarrow e^-\gamma j (j)$ after the preselection in Eq.~\eqref{eq:cut2}.} \label{fig:dis}
\end{figure}

In the signal events, the photons arise from the decay of the ALP and thus exhibit a peak around $m_a$ in the invariant mass distribution. However,  the photons for the backgrounds come from the radiation of the electron and quark, thus the peak position of the invariant mass distribution from the photon pair is determined by the transverse momentum of the photons; see Fig.~\ref{fig:dis} for the normalized invariant mass distributions of the photon pair from the signal and backgrounds. Owing to the typical decay width of the ALP is much smaller than the resolution of di-photon invariant mass at the EIC~\cite{AbdulKhalek:2021gbh},
we further require the invariant mass of the first two leading photons within the mass window,
 \begin{align}
|m_{\gamma\gamma} -m_a |<5 ~\GeV. \label{cut:photon}
\end{align}
We also checked that the other kinematic observable (e.g. $p_T^{e,j,\gamma}$) can not improve the cut efficiency  significantly in this process. It arises from the fact that  both the signal and backgrounds share a similar topology of the Feynman diagram, as a result, both of them should exhibit a similar $p_T^{e,j}$ distribution.  On the other hand, the photons share the energy from the ALP in the signal process, thus  the $p_T^\gamma$ distribution should peak around $\sim m_a/2$, while the $p_T^\gamma$ for the backgrounds tend to be a soft spectrum  because of the cross sections will be enhanced by the soft and/or collinear singularity in that phase space region. However, the $p_T^\gamma$ information could be correlated to the $m_{\gamma\gamma}$, therefore,  the additional cut on the $p_T^\gamma$ can not improve the sensitivity to probe the ALP.  We show the cut efficiencies with few benchmark ALP mass after the kinematic cuts from Eqs.~\eqref{eq:cut2} and ~\eqref{cut:photon} in Table~\ref{TAB:cuts}. 
It shows that the cut efficiencies of the reducible backgrounds ($e\gamma j$ and $e\gamma jj$) are smaller than irreducible backgrounds ($e\gamma\gamma j$ and $e\gamma\gamma jj$). But, the reducible backgrounds would be still dominated after we taking into account the production rates of those processes.
Owing to the kinematic cuts in Eq.~\eqref{eq:cut2} could be satisfied automatically for a heavy ALP, the cut efficiency for the signal tends to  be a constant when $m_a>40~{\rm GeV}$.

\begin{table}
\small
\centering
\begin{tabular}{|c|c|c|c|c|c|c|}
\hline
$m_a (\GeV )$&11&21&31&41&51&61\\
\hline
$\epsilon(\text{signal})$& 0.394&0.542&0.712&0.746&0.703&0.742\\
\hline
$\epsilon(e^-\gamma\gamma j)$ &0.066&0.100&0.046&0.011& $1.1\ast10^{-3}$&$1.7\ast10^{-4}$\\
\hline
$\epsilon(e^-\gamma\gamma jj)$ &0.070&0.098&0.033&0.005&$1.9\ast10^{-4}$&$3.9\ast10^{-5}$\\
\hline
$\epsilon(e^-\gamma j)$ & 0.0004 & 0.0007 & 0.0005 & 0.0003 & $0.0002$ & $0.0001$ \\
\hline 
$\epsilon(e^-\gamma j j)$ & 0.0003 & 0.0004 & 0.0003 & 0.0001 & $6.6\ast 10^{-5}$ & $5.7\ast 10^{-5}$ \\
\hline
\end{tabular}
\caption{Cut efficiencies for the signal process with the benchmark ALP mass $m_a$ and the corresponding SM backgrounds at the EIC.} \label{TAB:cuts}
\end{table}

\begin{figure}
\centering
\includegraphics[scale=0.6]{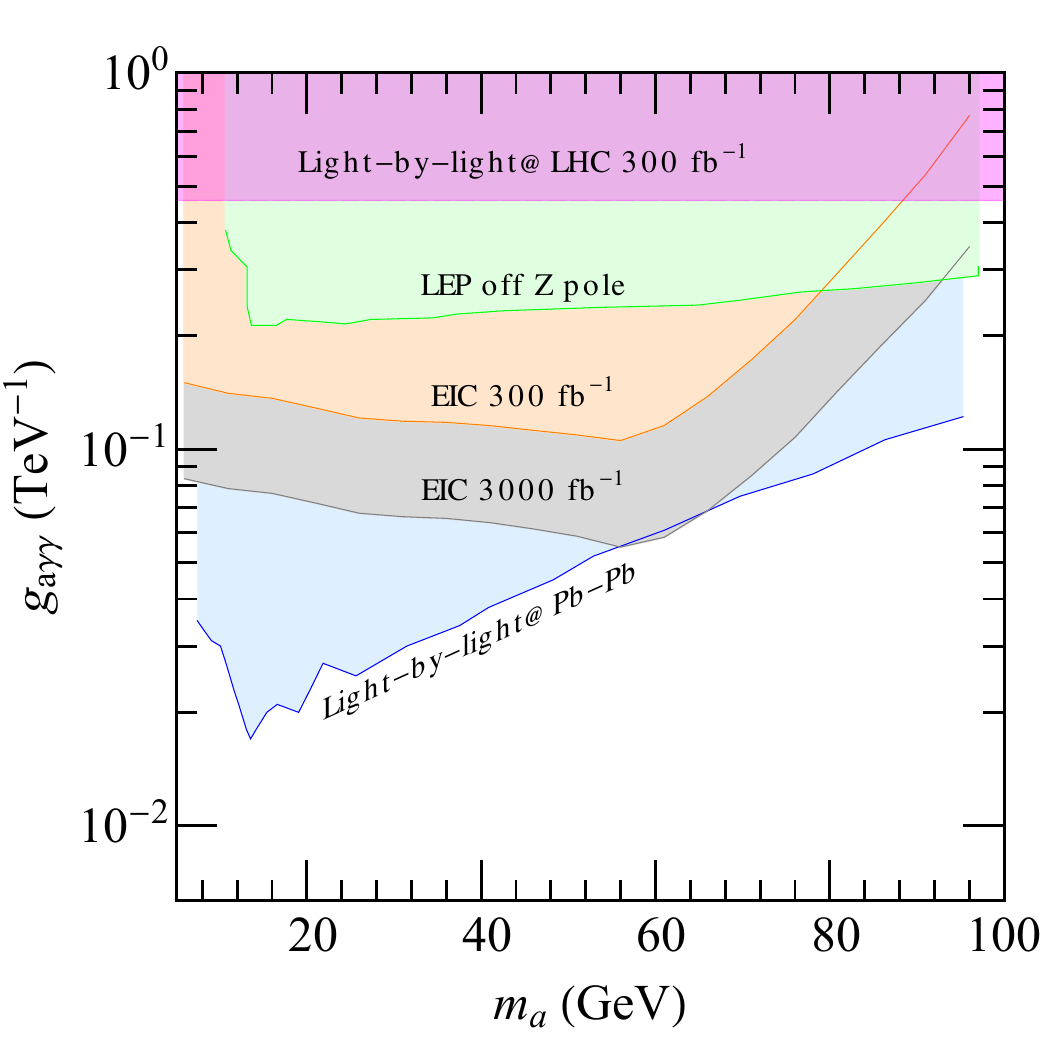}
\caption{Upper limit on the effective coupling $g_{a\gamma\gamma}$ from the EIC with proton beam (orange region) and Pb beam (gray region), Light-by-Light scattering with p-p collision with center-of-mass energy $\sqrt{s}=13~{\rm TeV}$ (pink region, projected sensitivity) and Pb-Pb collision with $\sqrt{s}=5.02~{\rm TeV}$ (blue region), off $Z$-pole at  the LEP (green region) and $Z\to \gamma a$ (yellow region). The excluded regions from other colliders are extracted from Ref.~\cite{Baldenegro:2018hng}. } 
\label{FIG:com}
\end{figure}

Equipped with the signal and backgrounds production cross sections and the collider simulation efficiencies, the upper limit on the effective coupling $g_{a\gamma\gamma}$ at $2\sigma$ confidence level can be obtained in terms of~\cite{Cowan:2010js}
\begin{align}
\sqrt{-2\left[n_b \ln\left(\frac{n_s+n_b}{n_b}\right)-n_s\right]} =2,
\label{eq:excl}
\end{align}
where $n_s$ and $n_b$ are the numbers of the signal and background events, respectively. Given the integrated luminosity of 300 fb$^{-1}$, the upper limit on the effective coupling $g_{a\gamma\gamma}$ is presented in Fig.~\ref{FIG:com} (orange region).  It shows that the $g_{a\gamma\gamma}$ could be constrained to be $\sim 0.2$ TeV$^{-1}$ at $2\sigma$  confidence level by assuming null result of directly searching of  ALP with $m_a<40~{\rm GeV}$ at the EIC.
This result could be further improved if we consider the nucleus beam at the EIC, because of the cross section could be enhanced by the atomic weight $A=Z+N$, where $Z$ and $N$ are the numbers of proton and neutron in nucleus~\cite{Davoudiasl:2021mjy}.  To roughly estimate the upper limit of the $g_{a\gamma\gamma}$ from nucleus beam, we use lead (Pb) as an example. The cross sections of signal and backgrounds from Pb beam could be obtained by properly rescaling the cross sections at electron-proton collision, i.e. $\sigma_{e^-{\rm Pb}}\simeq A\sigma_{e^-{\rm P}}$, with $A=208$ for Pb beam~\footnote{Note that the photon fusion scattering in this work is different from the Light-by-Light scattering in Ref.~\cite{Davoudiasl:2021mjy}. 
The photon flux for our case is proportional to the atom number, while it depends on the charge of the beam for the Light-by-Light scattering.}. Although the mixing of protons and neutrons in nucleus will change the total density for up and down quarks in the nuclear PDFs relative to the proton ones, the total effects will mildly modify the total cross sections and kinematic distributions~\cite{Li:2020rqj}, which do not alter the conclusion remarkably in this study. Under this approximation, we obtain 
the effective coupling $g_{a\gamma\gamma}$ from Pb beam could be improved by few times compared to the proton beam; see the gray region of Fig.~\ref{FIG:com} .

\section{Summary and discussion}
As we discussed in Sec.~\ref{sec:intr}, the tens of GeV ALP has been widely searched by other colliders.  It includes the measurements of tri-photon  on and off the $Z$-pole $(e^+e^-\to 3\gamma)$ at the LEP~\cite{DELPHI:1999fgt,L3:1994shn}, searches for the same final states at the LHC~\cite{Jaeckel:2015jla} and the Light-by-Light scattering in the heavy-ion collisions at the LHC energy and pp collisions at the LHC~\cite{ATLAS:2017fur,CMS:2018erd,Baldenegro:2018hng}.  However, the bound of the $g_{a\gamma\gamma}$ from some above measurements are depending on the assumption of the $aZ\gamma$ interaction. 
It arises from the fact that 
both the couplings could be generated from dimension-5 operators $g^{\prime 2}C_{BB}/\Lambda aB_{\mu\nu}\widetilde{B}^{\mu\nu}$ and $g^2C_{WW}/\Lambda aW_{\mu\nu}^A\widetilde{W}^{A,\mu\nu}$, where $B_{\mu\nu}$ and $W_{\mu\nu}^A$ are the field strength tensors of $U(1)_Y$ and $SU(2)_L$, $g^\prime$ and $g$ are corresponding gauge couplings. The effective coupling strengths of $a\gamma\gamma$ and $aZ\gamma$ are $g_{a\gamma\gamma}\sim C_{WW}+C_{BB}$ and  $g_{aZ\gamma}\sim c_W^2C_{WW}-s_W^2C_{BB}$, respectively. Therefore, the couplings $g_{a\gamma\gamma}$ and $g_{aZ\gamma}$ can be related to each other when we consider one operator at a time.

We show the comparison of probing the $g_{a\gamma\gamma}$ at the EIC with other measurements in Fig.~\ref{FIG:com}.
It is evident that the $Z$-pole measurement $Z\to \gamma a$ (yellow region), as compared to the other processes, yields the strongest constraint on the value of $g_{a\gamma\gamma}$. But this conclusion is only available when we consider one operator in the analysis.
We also notice that the expected limit from the EIC with proton beam (orange region) could be stronger than the off $Z$-pole measurement (green region) at the LEP and the Light-by-Light scattering at the LHC (pink region),  
while the expected limit from the Pb beam at the EIC (gray region) could be comparable to the measurement at the the Pb-Pb collision (blue region). 
Finally, we emphasize that the photon fusion production at the EIC is complementary to the other processes in the measurement of the ALP-photon coupling.

\medskip
\begin{acknowledgments}
The authors thanks Xiaohui Liu and Hao Zhang for helpful discussion. The work of Y. Liu is supported in part by the National Science Foundation of China under Grand No. 11805013, 12075257, BY is supported by the U.S. Department of Energy, Office of Science,
Office of Nuclear Physics, under Contract DE-AC52-06NA25396 through the LANL/LDRD Program, as well as the TMD topical collaboration for nuclear theory and IHEP under Contract No. E25153U1.
\end{acknowledgments}

\medskip
\bibliographystyle{apsrev}
\bibliography{reference}

\end{document}